\documentclass[aps,pra,twocolumn,showpacs,showkeys,groupedaddress]{revtex4}

\usepackage{amsmath,amssymb,amsthm,graphicx}
\usepackage{subfigure,psfrag}

\newcommand{\Tr}{\mathrm{Tr}}
\newcommand{\sumhight}{{\displaystyle\vphantom{\sum_{i=1}^m}}}

\newcommand{\bra}[1]{\langle #1 |}
\newcommand{\ket}[1]{| #1 \rangle}
\newcommand{\bracket}[2]{\langle #1 | #2 \rangle}

\def\argmax{\mathop{\rm argmax}}

\newtheorem{theorem}{Theorem}
\newtheorem{lemma}{Lemma}
\newtheorem{corollary}{Corollary}

\begin{document}

\preprint{Draft}

\title{Quantum Detection with Uncertain States}


\author{Noam Elron}
\email[]{nelron@tx.technion.ac.il}
\affiliation{Department of Electrical Engineering \\
             Technion - Israel Institute of Technology \\
             Technion City, Haifa 32000, Israel}

\author{Yonina C. Eldar}
\email[]{yonina@ee.technion.ac.il}
\affiliation{Department of Electrical Engineering \\
             Technion - Israel Institute of Technology \\
             Technion City, Haifa 32000, Israel}


\date{\today}

\begin{abstract}
We address the problem of distinguishing among a finite collection
of quantum states, when the states are not entirely known.
For completely specified states, necessary and sufficient
conditions on a quantum measurement minimizing the probability of
a detection error have been derived.
In this work, we assume that each of the states in our collection
is a mixture of a known state and an unknown state.
We investigate two criteria for optimality.
The first is minimization of the \emph{worst-case} probability of
a detection error.
For the second we assume a probability distribution on the unknown
states, and minimize of the \emph{expected} probability of a detection
error.

We find that under both criteria, the optimal detectors are equivalent
to the optimal detectors of an ``effective ensemble''.
In the worst-case, the effective ensemble is comprised of the known states
with altered prior probabilities, and in the average case it is made up of
altered states with the original prior probabilities.
\end{abstract}


\maketitle

\section{Introduction}

Quantum detection refers to the retrieval of classical information
encoded in a quantum-mechanical medium.
Representing the information as one of $m$ possible messages, it is
assumed that this medium has been prepared in a quantum state drawn
from a collection of $m$ known states, each associated with one of
the messages.
The medium is then subjected to a quantum measurement, in order to
determine the prepared state.
If the quantum states are not mutually orthogonal, then no measurement
will distinguish perfectly between them.
One, then, seeks a measurement scheme (detector), which optimally
discriminates between the states in some sense.
A popular criterion of optimality is minimization of the probability
of a detection error.

Possible applications for distinguishing between quantum states are
digital communication via a quantum channel, or the output module
of a quantum computer \cite{NielsenChuang}.
In theoretical quantum computation, the possible outcomes of a
calculation are normally mutually orthogonal, making the
discrimination between the results trivial.
In this paper, however, we address questions of imperfections in the
setup, making the results relevant to the implementation of working
quantum computers.

We consider an ensemble of quantum states, consisting of $m$ density
operators $\{\rho_i\}_{i=1}^m$ on an $n$-dimensional complex Hilbert
space $\mathcal{H}$, with prior probabilities $\{p_i\}_{i=1}^m$.
A density operator $\rho$ is a positive semidefinite (PSD) Hermitian
operator with $\Tr(\rho)=1$; we write $\rho \geq 0$ to indicate that
$\rho$ is PSD.
For our \emph{measurement}, we consider general positive operator-valued
measures \cite{Peres:BOOK,Peres:Neumark}, consisting of $m$ PSD Hermitian
operators $\{\Pi_i\}_{i=1}^m$ that form a resolution on the identity on
$\mathcal{H}$.

For a completely specified state set $\{\rho_i\}_{i=1}^m$, necessary
and sufficient conditions for an optimal measurement, which minimizes
the probability of a detection error, have been derived
\cite{Holevo1,YKL1,YoninaMegretskiVerghese1}.
Explicit solutions to the problem are known in some particular cases
\cite{Helstrom1,CBH,OBH,BKMH,YoninaForney1}, including ensembles
obeying a large class of symmetries \cite{YoninaMegretskiVerghese2}.

For arbitrary state sets, the problem of finding the optimal measurement
can be cast as a semidefinite programme (SDP) \cite{YoninaMegretskiVerghese1},
which is a tractable convex optimization problem \cite{Boyd&Vanden}.
By exploiting the many well-known algorithms for solving
SDPs \cite{VandenBoyd:SDP,NesterovNemirovsky}, the optimal measurement can
be computed very efficiently in polynomial time within any desired
accuracy.

As with most physical systems, typically, one does not have full
knowledge of the parameters.
When applying the measurement, the states $\rho_i$ are often unknown
to a certain extent, whether due to degradation (noise, decoherence
\cite{sep-qm-decoherence}) in the quantum medium, or to imperfect
preparation.
In this paper we investigate the effects of uncertainty in the states
$\rho_i$ on the optimal measurement.

To model the uncertainty we assume that each state $\rho_i$ is a mixture
of a known state and an unknown state
\begin{equation} \label{uncertainty}
\rho_i \triangleq q_i \rho _i^0 + (1-q_i) \rho_i^1
\end{equation}
where the states $\rho_i^0$ and $0 \leq q_i \leq 1$ are
known, and the operators $\rho_i^1$ are completely unspecified,
except for being valid quantum states.
The parameters $q_i$ serve as a bound on the amount of mixing
of each state.

A different detection strategy, known as \emph{unambiguous
detection} \cite{Ivanovic,YoninaUnambig,YoninaStojnicHassibi}, is to design
a measurement of order $m+1$, where the extra answer stands for an inconclusive
result.
If the measurement returns an answer, then it is correct with probability 1.
The goal is to design the measurement, so that the probability of an
inconclusive result is small.
When the states that are to be detected are uncertain as in \eqref{uncertainty},
ensuring perfect detection of a state is impossible.
For this reason we choose not to pursue this strategy.

Detection of uncertain states has so far been addressed in the
special case, where the quantum medium is the free space channel
and the known states $\rho_i^0$ are coherent states, i.e. pure
states $\ket{\alpha_i}$, each characterized by a complex number
$\alpha_i$ \cite{VilnrotterLau1}.
Vilnrotter and Lau \cite{VilnrotterLau2} model thermal noise as a
probability distribution over a finite subset $\{\alpha_j\}$ of the
complex plane, and then find the optimal measurement for states mixed
according to this distribution, i.e. $\rho_i^{\text{eff}} = \sum_j q_j \ket{\alpha_i+\alpha_j}\bra{\alpha_i+\alpha_j}$.
Concha and Poor \cite{ConchaPoor} use Ohya's model \cite{Ohya} of
quantum channels to model thermal noise in a multiaccess quantum
free space channel.
Both works rely heavily on the simple parameterization of coherent
states.
This type of parametric averaging may not suit all quantum systems.
In addition, the method we propose in this work assumes far less
knowledge (only the probability $q_0$ associated with $\alpha_j=0$).

The measurement minimizing the probability of a detection error
depends, in general, on the states $\rho_i$.
Therefore, if the states are not known exactly, then the optimal
measurement cannot be determined.
Here, we develop two approaches to detection in the presence of
state uncertainty.
The first strategy is motivated by the recent theory of robust
optimization \cite{BenTalNem1,EGOL,BTEGN}, in which the worst-case
solution is optimized.
Adapting this method to our particular context, we consider
maximizing the smallest possible probability of correct detection,
over all quantum states of the form \eqref{uncertainty}.
Robust semidefinite programming has already been introduced to
the domain of quantum information in the context of entanglement
witnesses \cite{FBRV}.
Our second strategy is to define a probability distribution over
the region of uncertainty, and then maximize the average probability
of correct detection.
This strategy is conceptually similar to those used in
\cite{VilnrotterLau2,ConchaPoor}, but does not rely on a particular
quantum system.

In Section \ref{formulation} we present the problem in detail and
state some known results.
Section \ref{WCdet} is an analysis of the worst-case approach to
minimal detection error, and in Section \ref{AVGdet} we address
the question of optimal detection on average.
In both cases we find that the optimal measurement for the uncertain
ensemble is equivalent to an optimal measurement for an ``effective
ensemble''.
The effective ensemble for worst-case detection is comprised of the
known states $\rho_i^0$ with altered prior probabilities, whereby a
bias towards the states which are more certain is introduced.
In the average case, it is made up of the states $q_i \rho_i^0 + \frac{1-q_i}{n} I$,
with the original prior probabilities.
We show explicitly that averaging over the region of uncertainty is
equivalent to choosing $\rho_i^1 = \frac{1}{n} I$, where $I$ is the
identity on $\mathcal{H}$.
The quantum state $\frac{1}{n} I$, known as the \emph{maximally mixed}
state, regularly serves in quantum mechanics to represent a complete
lack of knowledge.

For both strategies we address the special case where the uncertainty
bounds are uniform in $i$ ($q_i=q$).
We find that for high values of $q$ the worst-case optimal measurement
coincides with the optimal nominal measurement.
Also, in the equiprobable case, the optimal average measurement is
identical with the optimal nominal measurement for any $q>0$.

Section \ref{numerical} contains the results and analysis of numerical
simulations of several examples.
We compare the characteristics and performance of the two approaches.

\section{Formulation of the Detection Setup}
\label{formulation}

Assume that a quantum channel is prepared in a quantum state drawn
from a finite collection of quantum states.
The quantum states are represented by a set of $m$ PSD Hermitian
density operators $\{\rho_i\}_{i=1}^m$ on an $n$-dimensional
complex Hilbert space $\mathcal{H}$.
The states $\rho_i$, however, are not entirely known at the receiver,
whose state of knowledge is defined by \eqref{uncertainty}.
The receiver performs a quantum measurement $\Pi$, comprising $m$ PSD
Hermitian operators $\Pi = \{\Pi_i\}_{i=1}^m$ on $\mathcal{H}$, in order
to determine which of the messages was sent.

We assume without loss of generality that the eigenvectors of the known
density operators $\{\rho_i^0\}_{i=1}^m$ span \footnote{Otherwise, we can
transform the problem to a problem equivalent to the one considered in
this paper, by reformulating the problem on the subspace spanned by the
eigenvectors of $\{\rho_i^0\}_{i=1}^m$.} $\mathcal{H}$ (in this case,
the eigenvectors of $q_i \rho_i^0 + (1-q_i) \rho_i^1$ also span
$\mathcal{H}$). Under this assumption, the measurement operators $\Pi_i$
must satisfy
\begin{equation} \label{OpSum}
\Pi_i \geq 0 , \quad \sum_{i=1}^m \Pi_i = I
\end{equation}
where $I$ is the identity on $\mathcal{H}$, in order to
be a valid measurement.
We shall denote the set of all $m$ order POVMs $\Pi = \{\Pi_i\}_{i=1}^m$ as
\begin{equation}
B \triangleq \left\{ \Pi \; \Bigg| \; \Pi_i \geq 0 , \quad \sum_{i=1}^m \Pi_i = I \right\}
\end{equation}

Given that the transmitted state is $\rho_j$, the probability of
correctly detecting the state using measurement $\Pi=\{\Pi_i\}_{i=1}^m$
is $\Tr(\Pi_j \rho_j)$.
Therefore, the probability of correct detection is given by
\begin{equation}
P_d = \sum_{i=1}^m p_i \Tr(\Pi_i \rho_i)
\end{equation}
where $p_i > 0$ is the a-priori probability of $\rho_i$,
with $\sum_i p_i = 1$.

When the states are known, we may seek the measurement $\Pi \in B$
that minimizes the probability of detection error, or equivalently,
maximizes the probability of correct detection.
This can be expressed in the form of the optimization problem
\begin{align} \label{aimProb}
& \max_{\Pi_i} \left\{ \sum_{i=1}^m p_i \Tr (\Pi_i \rho_i) \right\} \\
& \qquad \text{s.t.} \begin{cases}
            \sumhight \Pi_i \geq 0 & \text{(a)} \\
            {\displaystyle\sum_{i=1}^m \Pi_i = I} \qquad & \text{(b)}
                     \end{cases} \nonumber
\end{align}
If $q_i = 1$ for all $i$, so that $\rho_i$ is completely
specified, then it was shown in \cite{Holevo1,YKL1,YoninaMegretskiVerghese1}
that a measurement $\{\hat{\Pi}_i\}_{i=1}^m \in B$ solves \eqref{aimProb}
if and only if there exists an operator $\hat{U}$ such that for all
$1 \leq i \leq m$
\begin{equation} \label{OrigN&S}
\begin{array}{c}
{\displaystyle\vphantom{\sum}} \hat{U} \geq p_i \rho_i \\
(\hat{U} - p_i \rho_i) \hat{\Pi}_i = 0 \end{array}
\end{equation}
(by the notation $X \geq Y$ we mean that $X-Y$ is PSD).
Throughout this paper we shall denote the measurement which solves
problem \eqref{aimProb} by $\Pi^{\rm Nom} = \hat{\Pi}(\rho_i,p_i)$.

In general, the optimal measurement will depend on the states
$\rho_i$.
Because in our formulation they are unknown, new criteria for
optimality must be defined.
Before doing so, we present a measurement, which is independent
of the states $\rho_i$, and will thus provide a lower bound on
the optimal probability of correct detection, regardless of the
criterion used.
The following measurement is dependent solely on the prior probability
distribution: for all $i$ choose
\begin{equation} \label{guess}
\bar{\Pi}_i = \begin{cases}
\frac{1}{m_{\max}} I, \qquad & p_i = p_{\max} \\
0,                           & p_i < p_{\max}
\end{cases}
\end{equation}
where $m_{\max} \leq m$ stands for the number of states
with prior probability $p_{\max}$.
Using this measurement, the probability of correct detection is
\begin{equation}
\bar{P}_d = \sum_{i=1}^m p_i \Tr (\Pi_i \rho_i)
= \sum_{j=1}^{m_{\max}} \frac{p_{\max}}{m_{\max}} \Tr (\rho_j) =  p_{\max}
\end{equation}
This detector is effectively an unbiased guess from the
subset of messages with maximal prior probability.

Thus, probability of correct detection equal to $p_{\max}$ can always
be achieved.
We seek a measurement that under conditions of uncertainty can perform
better.
In the following two sections, we derive necessary and sufficient
conditions for optimal measurements, using two different criteria.
These criteria refer to the probability of correct detection, while
specifying a certain point in the region of uncertainty.
In Section \ref{WCdet} we present the optimal worst-case measurement,
and in Section \ref{AVGdet} we propose an optimal expected measurement
under the assumption of a probability distribution for the unknown
states $\rho_i^1$.

\section{Optimal Worst-Case Detection}
\label{WCdet}

In a worst-case or \emph{robust} approach, we first find the point in
the region of uncertainty that would, for a given measurement, yield
the poorest outcome.
We then solve the ``original'' optimization problem for this point.
This criterion serves to assure that the probability of correct detection
obtained using the optimal detector will not be lower than a certain value
(the optimal value).
An uncertainty model such as ours, that assumes very little prior knowledge
(only the bounds $q_i$), can be regarded as possessing a worst-case quality -
adding prior knowledge will surely improve the performance of the optimal
measurement.
This observation makes this specific criterion especially interesting.

Using the uncertainty model \eqref{uncertainty} and problem \eqref{aimProb},
the worst-case measurement is the solution to
\begin{align} \label{BigProb}
&\max_{\Pi \in B} \min_{\rho_i^1} \left\{ \sum_{i=1}^m p_i \Tr
      [\Pi_i (q_i \rho_i^0 + (1-q_i) \rho_i^1)] \right\} \\
& \quad \text{s.t.} \begin{cases}
        \rho_i^1 \geq 0 \qquad \qquad & \text{(a)} \\
        \Tr(\rho_i^1) = 1 & \text{(b)}
                    \end{cases} \nonumber
\end{align}
where the constraints represent valid measurements,
and the region of uncertainty.
We shall denote the optimal worst-case probability of correct
detection as $P_d^{\rm WC} (\vec{q})$.

We begin by proving our first result, stated in Theorem \ref{thm1}.
We then explore in further detail the special case in which $q_i = q$ for
all $i$.

\begin{theorem} \label{thm1}
Let $\{\rho_i = q_i \rho_i^0 + (1-q_i) \rho_i^1\}_{i=1}^m$ be a set of
quantum states, where $\rho_i^0$ and $0 \leq q_i \leq 1$ are known and
the states $\rho_i^1$ are unknown.
Each state has prior probability $p_i$.
Denote $\eta(\vec{q}) \triangleq \sum_{i=1}^m p_i q_i$, and define
``effective probabilities''
\[
\tilde{p_i} = \frac{p_i q_i}{\eta(\vec{q})}
\]
Denote the optimal measurement on the ``effective ensemble''
(the states $\{\rho_i^0\}_{i=1}^m$ with prior probabilities $\tilde{p_i}$)
\[
\tilde{\Pi} = \hat{\Pi}(\rho_i^0,\tilde{p}_i) =
    \argmax_{\Pi \in B} \left\{ \sum_{i=1}^m \tilde{p}_i \Tr (\Pi_i \rho_i^0) \right\}
\]
and the probability of correct detection achieved by $\tilde{\Pi}$
on the effective ensemble
\[
\tilde{P}_d = \sum_{i=1}^m \tilde{p}_i \Tr (\tilde{\Pi}_i \rho_i^0).
\]
The quantum measurement $\Pi^{\rm WC}=\{\Pi_i^{\rm WC}\}_{i=1}^m$ that minimizes
the worst-case probability of a detection error is
\[
\Pi^{\rm WC} (\vec{q}) = \begin{cases}
               \tilde{\Pi} \qquad &\eta(\vec{q}) \tilde{P}_d > p_{\max} \\
               \bar{\Pi}          &\eta(\vec{q}) \tilde{P}_d \leq p_{\max} \end{cases}
\]
where $\bar{\Pi}$ is defined in \eqref{guess}.

The worst-case probability of correct detection is
\[
P_d^{\rm WC} (\vec{q}) = \begin{cases}
                         \eta (\vec{q}) \tilde{P}_d \qquad & \Pi^{\rm WC}=\tilde{\Pi} \\
                         p_{\max}                          & \Pi^{\rm WC}=\bar{\Pi} \end{cases}
\]
\end{theorem}

\begin{proof}[\textup{\textbf{Proof:}}]
The internal minimization in problem \eqref{BigProb} can be written as
\begin{align} \label{innerProb}
& \min_{\rho_i^1} \left\{ \sum_{i=1}^m p_i [q_i \Tr (\Pi_i \rho_i^0) +
        (1-q_i) \Tr (\Pi_i \rho_i^1) ] \right\} \\
& \quad \text{s.t.} \begin{cases}
        \rho_i^1 \geq 0 \qquad \qquad & \text{(a)} \\
        \Tr(\rho_i^1) = 1 & \text{(b)}
                    \end{cases} \nonumber
\end{align}
where the parameters $\{\Pi_i\}_{i=1}^m$ are all PSD.

Since the objective function is dependant on $\rho_i^1$ only through
the expression on the right, and since it is also separable in $i$
(the objective is additive and the constraints independent), the
optimization is reduced to solving $m$ cases of the form
\begin{align} \label{sepProb}
& \min_{\rho} \Tr (\Pi \rho) \\
& \quad \text{s.t.} \begin{cases}
        \rho \geq 0 \quad \quad & \text{(a)} \\
        \Tr(\rho) = 1 & \text{(b)}
                    \end{cases} \nonumber
\end{align}
By writing $\rho$ as a convex combination of pure states
\begin{equation}
\rho = \sum_{j=1}^n g_j \ket{\mu_j} \bra{\mu_j}
\end{equation}
the problem \eqref{sepProb} can be recast as
\begin{align} \label{sepProb2}
& \min_{g_j,\ket{\mu_j} } \sum_{j=1}^n g_j \bra{\mu_j} \Pi \ket{\mu_j} \\
& \quad \text{s.t.} \begin{cases}
        \sum_j g_j = 1 \quad \quad & \text{(a)} \\
        \bracket{\mu_j}{\mu_j} = 1 & \text{(b)}
                    \end{cases} \nonumber
\end{align}
For each $j$, the minimal value of $\bra{\mu_j} \Pi \ket{\mu_j}$
is the minimal eigenvalue of $\Pi$, (denoted $\lambda_{\Pi}^{\min}$)
and is achieved for a state $\ket{\hat{\mu_j}}$, which lies in the
corresponding eigenspace.
The optimal value of \eqref{sepProb2} is therefore
$\sum_j g_j \lambda_{\Pi}^{\min} = \lambda_{\Pi}^{\min}$ and is achieved
for a state $\hat{\rho}$ whose range-space lies entirely in the associated
eigenspace.

Using this result in \eqref{innerProb}, the original problem \eqref{BigProb}
is equivalent to
\begin{equation} \label{BigProb2}
P_d^{\rm WC} (\vec{q}) = \max_{\Pi \in B} \left\{ \sum_{i=1}^m p_i
      [ q_i \Tr (\Pi_i \rho_i^0) + (1-q_i) \lambda_{\Pi_i}^{\min} ] \right\}
\end{equation}

By utilizing the fact that the minimal eigenvalue of a Hermitian operator
$\Pi$ can be written as the solution to
\begin{equation} \label{minEigVal}
\lambda_{\Pi}^{\mathrm{min}} = \left\{ \begin{split}
& \max_{\lambda \in \mathcal{R}} \lambda \\
& \quad \text{s.t.} \quad \Pi \geq \lambda I
\end{split} \right.
\end{equation}
problem \eqref{BigProb2} becomes
\begin{align} \label{AnalProb}
P_d^{\rm WC} (\vec{q}) = & \max_{\Pi_i,\lambda_i} \left\{ \sum_{i=1}^m p_i
        [q_i \Tr (\Pi_i \rho_i^0) + (1-q_i) \lambda_i ] \right\} \\
& \quad \text{s.t.} \begin{cases}
        \sumhight \Pi_i \geq \lambda_i I \geq 0 \quad \quad & \text{(a)} \\
        {\displaystyle\sum_{i=1}^m \Pi_i = I} & \text{(b)}
                    \end{cases} \nonumber
\end{align}

The objective function in problem \eqref{AnalProb} is linear and the
constraint are all linear matrix equalities and inequalities, making it
a convex problem.
For convex optimization problems which are strictly feasible (i.e. the
feasibility set has a non-empty relative interior), necessary and sufficient
conditions for optimality are given by the \emph{Karush-Kuhn-Tucker}
(KKT) conditions \cite{Boyd&Vanden}.
In Appendix \ref{pndx1} we derive the KKT conditions for the specific problem
\eqref{AnalProb}:
A measurement $\tilde{\Pi} \in B$ with bounds $\tilde{\lambda}_i \geq 0$ are optimal
if there exists a Hermitian operator $\tilde{U}$ satisfying
\begin{gather}
\tilde{\Pi}_i \geq \tilde{\lambda}_i I \nonumber \\
\tilde{U} \geq p_i q_i \rho_i^0 \nonumber \\
\Tr (\tilde{U}) \geq p_i \label{N&S1}\\
(\tilde{U} - p_i q_i \rho_i^0) (\tilde{\Pi}_i - \tilde{\lambda}_i I) = 0 \nonumber \\
(\Tr(\tilde{U}) - p_i ) \tilde{\lambda}_i = 0 \nonumber
\end{gather}

Moreover, from Lagrange Duality we know that the optimal values of
$\tilde{\Pi}_i$, $\tilde{\lambda}_i$ and $\tilde{U}$
obey the relation
\begin{equation} \label{duality}
P_d^{\rm WC} (\vec{q}) = \Tr(\tilde{U})
\end{equation}

From \eqref{duality} we see that the aforementioned lower bound
$p_{\max}$ on the optimal probability of correct detection is
manifested in the requirement $\Tr(\tilde{U}) \geq p_{\max} \geq p_i$.
In \eqref{guess} we give a measurement which achieves $P_d = p_{\max}$,
and shall therefore continue our analysis of the worst-case under the
assumption $P_d^{\rm WC} (\vec{q}) > p_{\max}$.

With this extra demand in place, the last necessary condition in
\eqref{N&S1} can only be satisfied if for all $i$, the eigenvalue
bounds $\hat{\lambda}_i$ equal zero.
The necessary and sufficient conditions \eqref{N&S1} reduce
to (after ignoring the now redundant constraints)
\begin{equation} \label{N&S_reduced}
\begin{array}{c}
{\displaystyle\vphantom{\sum}} \tilde{U} \geq p_i q_i \rho_i^0\\
(\tilde{U} - p_i q_i \rho_i^0) \tilde{\Pi}_i = 0 \end{array}
\end{equation}

Denoting $\eta (\vec{q}) \triangleq \sum_i p_i q_i$, and
\begin{equation} \label{effectiveProbs}
\tilde{p}_i = \frac{p_i q_i}{\eta (\vec{q})}
\end{equation}
these conditions (with $(1/\eta)\tilde{U}$ in place of $\hat{U}$)
are identical to the necessary and sufficient conditions \eqref{OrigN&S},
for the known states $\rho_i^0$ with prior probabilities $\tilde{p}_i$.
\end{proof}

The larger $q_i$ is, the greater the ratio $\tilde{p}_i / p_i$ between
the ``effective'' prior probability and the real one.
When assuming $P_d^{\rm WC} (\vec{q}) > p_{\max}$, the optimal
worst-case measurement $\Pi^{\rm WC}$ is biased towards detecting
the states with less uncertainty.

Intuitively, with no knowledge at all about the states $\rho_i^1$,
all that can be inferred about the uncertain part of the ensemble
relies on the prior probabilities.
This is the reason that the uncertain ensemble is (in terms of
optimal detection) equivalent to an ensemble comprised of the
known states $\rho_i^0$ with altered prior probabilities.

In the extreme case where a state $\rho_j$ is completely unknown
($q_j = 0$), the optimal worst-case detector ignores it entirely
(because $\tilde{p}_j = 0$) and attempts to distinguish optimally
between the remaining $m-1$ states.

The next two corollaries give the probability of correct detection
when the known states $\rho_i^0$ are mutually orthogonal.

\begin{corollary} \label{cor1}
If $\rho_i^0$ are mutually orthogonal, and for all $i$, $q_i > 0$,
then the worst-case probability of correct detection is
\[
P_d^{\rm WC} (\vec{q}) = \max \left\{ p_{\max} , \eta (\vec{q}) \right\}
\]
\end{corollary}

\begin{proof}[\textup{\textbf{Proof:}}]
Provided that the effective prior probabilities satisfy $\tilde{p}_i > 0$,
when $\rho_i^0$ are mutually orthogonal they can always be correctly detected.
Therefore $\tilde{P}_d = 1$, and the corollary follows immediately from
Theorem \ref{thm1}.
\end{proof}

\begin{corollary} \label{cor2}
Denote by $I_0$ the index set of the states which are completely unknown,
i.e. have $q_i = 0$.
Denote $P_0 = P(i \in I_0) = \sum_{i \in I_0} p_i$ and the maximal prior
probability of a state from this subset $p_0 = \max\{p_i | i \in I_0 \}$.

If $\rho_i^0$ are mutually orthogonal, and $I_0 \neq \emptyset$, then the
worst-case probability of correct detection is
\[
P_d^{\rm WC} (\vec{q}) = \max \left\{ p_{\max} ,
     \eta (\vec{q}) \left( 1 - P_0 + p_0 \right) \right\}
\]
\end{corollary}

\begin{proof}[\textup{\textbf{Proof:}}]
The states $\rho_i^0$ with $\tilde{p}_i > 0$ are mutually orthogonal, and
can therefore be detected correctly.
In other words, when $i \notin I_0$, the conditional probability of correct
detection is $P(d|i \notin I_0)=1$.
If $i \in I_0$, then the state must be guessed from within this subset.
An optimal guess achieves $P(d|i \in I_0) = p_0 / P_0$.

All in all, the optimal measurement on the effective ensemble achieves
\begin{equation}
\begin{split}
\tilde{P}_d &= P(i \notin I_0)P(d|i \notin I_0) + P(i \in I_0)P(d|i \in I_0) \\
  & = (1-P_0) \cdot 1 + P_0 \frac{p_0}{P_0} = 1 - P_0 + p_0
\end{split}
\end{equation}
The corollary follows immediately from Theorem \ref{thm1}.
\end{proof}

Note that when there is one state with $q_i = 0$, we get $P_0 = p_0$.
The completely unknown state can be guessed by default.

\subsection{Worst-Case Detection with Uniform Uncertainty}

We now consider the special case in which the mixing bounds $q_i$
are uniform in $i$, i.e
\begin{equation} \label{uniformModel}
\rho_i = q \rho_i^0 + (1-q) \rho_i^1
\end{equation}

\begin{corollary} \label{UniformWC}
Denote the optimal nominal measurement $\Pi^{\rm Nom}=\hat{\Pi}(\rho_i^0,p_i)$,
and the nominal probability of correct detection $P_d^{\rm Nom}$.

When the uncertainty is uniform, as in \eqref{uniformModel}, then the
optimal worst-case measurement is
\[
\Pi^{\rm WC} (q) = \begin{cases}
\bar{\Pi},            & 0 \leq q \leq \dfrac{p_{\max}}{P_d^{\rm Nom}} \\
\Pi^{\rm Nom}, \qquad & \dfrac{p_{\max}}{P_d^{\rm Nom}} < q \leq 1 \end{cases}
\]
and the optimal probability of correct detection is
\[
P_d^{\rm WC} (q) = \max \big\{ p_{\max} , q P_d^{\rm Nom} \big\}
\]
\end{corollary}

\begin{proof}[\textup{\textbf{Proof:}}]
When the uncertainty is uniform for all $i$, the effective probabilities
defined in \eqref{effectiveProbs} are $\tilde{p}_i = p_i$.
The optimal measurement on the effective ensemble is therefore the one
which would have been optimal for the known states $\rho_i^0$ with the
original prior probabilities $p_i$, had there not been any uncertainty,
i.e. $\tilde{\Pi} = \Pi^{\rm Nom}$.

In addition,
\begin{equation}
\eta(q) = \sum_{i=1}^m p_i q = q
\end{equation}

The corollary then follows from Theorem \ref{thm1}.
\end{proof}

Corollary \ref{UniformWC} implies, that under uniform uncertainty with
a large value of $q$, the best course of action in terms of worst-case
performance is to ignore the uncertainty altogether.
Nonetheless, although it is achieved by the optimal nominal measurement,
the probability of correct detection itself is affected by the
uncertainty (see Section \ref{numerical}).

The complete symmetry in the uncertainty (total lack of knowledge about
$\rho_i^1$ and equal mixing) does not bias the optimal measurement in any
way, thus leaving it fixed with change in $q$, until the threshold is
reached.

\section{Optimal Average Detection}
\label{AVGdet}

Optimality in the worst-case does not grant good performance throughout
the region of uncertainty.
Also, as seen in the previous section, the optimal worst-case measurement
is sometimes quite pessimistic, altogether ignoring the input state in
favor of a guess.
An alternative course of action is to define a distribution of probability
over the region of uncertainty, thus enabling us to find a measurement,
which on average maximizes $P_d$.

Our model of uncertainty \eqref{uncertainty} assumes a complete lack
of knowledge about the states $\rho_i^1$.
This suggests two attributes of the probability distribution we shall
define:
\begin{enumerate}
\item The different unknown states are statistically independent of
      each other,
\item For each $i$, $\rho_i^1$ is distributed uniformly over the entire
      set of $n$-dimensional quantum states.
\end{enumerate}

A pure random state is equivalent to rotating an arbitrary state
using a random rotation $\ket{\mu} = U \ket{\mu_0}$.
Thus, the probability distribution of a ``uniformly distributed''
random pure state can be defined using the uniform measure on the group
of order $n$ rotation operators $SU(n)$, the well known Haar measure
\cite{Wootters1,Jones1}.
There does not, however, seem to be any natural uniform measure on
the set of mixed states \cite{Wootters1} (i.e. it is not simple to
provide rational arguments for the superiority of a given measure).
Many attempts to define such a distribution rely on \emph{product
measures} \cite{Hall1,ZS1}, whereby a random diagonal operator $G$,
distributed using a measure on the simplex of eigenvalues, is
rotated using a random unitary operator $U$, distributed using the
Haar measure
\begin{equation}
\rho_{\rm rand} = U G U^*
\end{equation}

Writing, the unknown states $\rho_i^1$ in similar fashion
\begin{equation} \label{CombPureStates}
\rho_i^1 = U_i G_i U_i^*
\end{equation}
we shall assume that each of the operators $U_i$ is a random
rotation operator distributed according to the Haar measure.
For all $i$, $U_i$ are statistically independent.
We assume nothing about the operators $G_i$ except that they are valid
quantum states ($G_i \geq 0$; $\Tr(G_i)=1$).

\begin{lemma} \label{Lemma1}
Let $\rho$ be a quantum state that has undergone a random rotation $U$
\[
\rho = UGU^*
\]
where $U$ is distributed using the Haar measure.
Then for any Hermitian operator $\Pi$, the expectation value of
$\Tr (\Pi \rho)$ is
\[
\langle \Tr(\Pi \rho) \rangle = \frac{1}{n} \Tr(\Pi)
\]
\end{lemma}

The proof of Lemma \ref{Lemma1} is given in Appendix \ref{pndx2}.

For a pure state $\rho$ this result can be intuitively understood.
The direction of the normalized vector is isotropically distributed, and so
the average measurement result of any operator on it is the average of the
possible outcomes (eigenvalues of the operator).
Lemma 1 states that this is also true for mixed states. \\*

We now wish to find the detector which maximizes the \emph{average}
probability of correct detection under the above probabilistic model,
i.e. we wish to find
\begin{equation} \label{AvgProb}
\max_{\Pi \in B} \left\{ E_{\rho_i^1} \left\{ \sum_{i=1}^m p_i \Tr
      [\Pi_i (q_i \rho_i^0 + (1-q_i) \rho_i^1)] \right\} \right\}
\end{equation}

The main result of this section is summarized in Theorem \ref{thm2}.

\begin{theorem} \label{thm2}
Given an ensemble of $m$ quantum states
$\{\rho_i = q_i \rho_i^0 + (1-q_i) U_i G_i U_i^*\}_{i=1}^m$,
where $\rho_i^0$ and $0 \leq q_i \leq 1$ are known, and $U_i$
are statistically independent random unitary matrices distributed
using the Haar measure, with prior probabilities $p_i$, obtaining
the quantum measurement $\Pi$ that minimizes the average probability
of a detection error is equivalent to obtaining the optimal measurement
for the ensemble $\{ q_i \rho_i^0 + \frac{1-q_i}{n} I \}_{i=1}^m$
with prior probabilities $p_i$.
\end{theorem}

\begin{proof}[\textup{\textbf{Proof:}}]
Using Lemma 1, the average probability of correct detection is given by
\begin{equation} \label{avgProbab}
\begin{split}
\langle P_d (\vec{q}) \rangle &= \sum_{i=1}^m p_i \bigg[ q_i \Tr (\Pi_i \rho_i^0)
                          + (1-q_i) \langle \Tr(\Pi_i \rho_i^1) \rangle \bigg] \\
&= \sum_{i=1}^m p_i \bigg[ q_i \Tr (\Pi_i \rho_i^0) + \frac{1-q_i}{n} \Tr(\Pi_i) \bigg] \\
&= \sum_{i=1}^m p_i \Tr [\Pi_i (q_i \rho_i^0 + \tfrac{1-q_i}{n} I)]
\end{split}
\end{equation}
Maximizing \eqref{avgProbab} requires finding the optimal detector
designed for states of the form $\rho_i = q_i \rho_i^0 + \frac{1-q_i}{n} I$
with prior probabilities $p_i$.
Using our established notation, $\Pi^{\rm Avg} = \hat{\Pi}(q_i \rho_i^0 + \frac{1-q_i}{n} I,p_i)$.
\end{proof}

Stating Theorem \ref{thm2} in other words, maximizing the average
probability of correct detection involves replacing the uncertainty with
maximally mixed states.
The maximization can done by directly solving the SDP \eqref{aimProb},
or by utilizing one of the closed form solutions, if applicable to the
new states $q_i \rho_i^0 + \frac{1-q_i}{n} I$.

The effective ensemble used in producing the optimal average measurement
has the original prior probabilities.
Therefore, the performance of the outcome of the optimization is bounded
below by the performance of the measurement in \eqref{guess}.
This implies, that the average performance is bounded below by $p_{\max}$,
but says nothing about the performance of this measurement in the worst
case (see numerical results in Subsection \ref{numB}).

\subsection{Average Detection with Uniform Uncertainty}

When $q_i = q$, the optimization problem that must be solved in
order to find the optimal average detector is
\begin{equation} \label{UniformAvgProb}
\max_{\Pi \in B} \left\{ q \sum_{i=1}^m p_i \Tr (\Pi_i \rho_i^0) +
        \frac{1-q}{n} \sum_{i=1}^m p_i \Tr (\Pi_i) \right\}
\end{equation}

Writing the measurement operators as $\Pi_i = \sigma_i \Psi_i$,
where $\Tr(\Psi_i)=1$, and $\sigma_i = \Tr(\Pi_i)$, we can regard
$\Psi_i$ as the ``geometry'' of the measurement operator, and
$\sigma_i$ as the ``relative importance'' of each operator.
Note that the posteriori probability of detecting the $i$-th message
\begin{equation}
P(i) = \sum_{j=1}^m p_j \Tr(\Pi_i \rho_j) = \sigma_i \Tr \left( \Psi_i \sum\nolimits_j p_j \rho_j \right)
\end{equation}
is proportional to $\sigma_i$.

Stating problem \eqref{UniformAvgProb} in these new terms, one gets
\begin{align} \label{UniformAvgProb2}
& \max_{\Psi_i,\sigma_i} \left\{ q \sum_{i=1}^m p_i \sigma_i \Tr (\Psi_i \rho_i^0) +
        \frac{1-q}{n} \sum_{i=1}^m p_i \sigma_i \right\} \\
        & \quad \text{s.t.} \begin{cases}
        \sumhight \Psi_i \geq  0 \qquad \Tr(\Psi_i) = 1 \qquad & \text{(a)} \\
        {\displaystyle\sum_{i=1}^m \sigma_i \Psi_i = I}        & \text{(b)}
                    \end{cases} \nonumber
\end{align}

The ``geometry'' of the measurement $\Psi_i$ are determined only through
the left-hand expression.
As $q$ grows smaller (more uncertainty), more importance in determining the
optimal measurement is given to the prior probability distribution, with less
regard to the states $\rho_i^0$ themselves.

When the different messages are equiprobable $p_i = \frac{1}{m}$, the right
hand term in \eqref{UniformAvgProb2} becomes a constant (with the value
$\frac{1-q}{mn}$).
The optimization problem reduces to finding the optimal measurement in the
nominal case.
Therefore, for all $q>0$ the optimal average measurement is identical to the
optimal nominal measurement.
Again we find that the optimal course of action is to simply ignore the
uncertainty when designing a measurement.

Numerical solutions of \eqref{aimProb} for the states
$q \rho_i^0 +\frac{1-q}{n} I$ reveal that when the prior probabilities
$p_i$ are not equal, many ensembles exhibit a certain value of $q$,
above which the optimal average measurement is equal to the optimal
nominal measurement.
This property, however, is not universal.

\section{Numerical Examples and Discussion}
\label{numerical}

\begin{table*}

\renewcommand{\thefootnote}{\fnsymbol{footnote}}

\caption{Summary of the optimal measurements using different criteria\label{thetable}}
\begin{ruledtabular}
\begin{tabular}{lllll}
\hfill\vline & \multicolumn{2}{c}{Different mixing bounds $q_i$} \vline
             & \multicolumn{2}{c}{Equal mixing bounds ($q_i = q$)} \\
\hline 

\begin{tabular}{l} Optimization\\criterion \end{tabular}
\hfill\vline & Measurement & Optimal $P_d$
\hfill\vline & \phantom{X} Measurement & Optimal $P_d$ \\
\hline \hline

\begin{tabular}{l} $\max P_d$ in\\nominal case \end{tabular}
\hfill\vline & $\Pi^{\rm Nom} \triangleq \hat{\Pi} (\rho_i^0,p_i)$
             & $P_d^{\rm Nom} \triangleq  \displaystyle\sum_{i=1}^m p_i \Tr(\Pi_i^{\rm Nom} \rho_i^0)$
\hfill\vline & \multicolumn{2}{c}{SIMILAR} \\
\hline

\begin{tabular}{l} $\max P_d$ in the\\worst case \end{tabular}
\hfill\vline & \phantom{X} or\footnote{One must choose between the two measurements, according to $P_d$ achieved by each.}
               $\;\;\; \begin{array}{l} \sumhight \hat{\Pi} (\rho_i^0,\tilde{p}_i) \\
                              \bar{\Pi} \end{array}$
             & $\begin{array}{l} \eta (\vec{q}) \displaystyle\sum_{i=1}^m \tilde{p}_i \Tr(\hat{\Pi}_i \rho_i^0) \\
               p_{\max} \end{array}$
\hfill\vline & \phantom{XXx} or \phantom{X} $\;\; \begin{array}{l} \sumhight \Pi^{\rm Nom} \\
                              \bar{\Pi} \end{array}$
             & $\begin{array}{l} \sumhight q P_d^{\rm Nom} \\ p_{\max} \end{array}$ \\
\hline

\begin{tabular}{l} $\max P_d$ on\\average \end{tabular}
\hfill\vline & $\begin{array}{l} \sumhight \Pi^{\rm Avg} \triangleq \hat{\Pi} (q_i \rho_i^0 + \frac{1-q_i}{n} I,p_i) \\ \phantom{X} \end{array}$
             & \begin{tabular}{l} $\sumhight P_d^{\rm Avg} (\vec{q})$\footnote{$P_d^{\rm Avg} (\vec{q})
                                  \triangleq \sum_{i=1}^m p_i \Tr [ \Pi_i^{\rm Avg}
                                  (q_i \rho_i^0 + \tfrac{1-q_i}{n} I) ].$} \\
                                  \phantom{X} \end{tabular}
\hfill\vline & \hspace{-17pt} \begin{tabular}{ll}                 & $\sumhight \Pi^{\rm Avg}$ \\
                    if $\forall i \; p_i = \tfrac{1}{m} \;\;\;\;$ & $\Pi^{\rm Nom}$ \end{tabular}
             & $\begin{array}{l} \sumhight P_d^{\mathrm{Avg}} (\vec{q}) \\
                                 q P_d^{\rm Nom} + \tfrac{1-q}{nm} \end{array}$
\end{tabular}
\end{ruledtabular}
\end{table*}

Table \ref{thetable} contains a summary of the results obtained in
the previous sections.
We have defined three criteria of optimality - nominal, worst-case
and average.
Throughout the following, we denote the three respective optimal
measurements as $\Pi^{\rm Nom}$, $\Pi^{\rm WC}$, and $\Pi^{\rm Avg}$.
We also refer to the measurement $\bar{\Pi}$, defined in \eqref{guess}.
The effective prior probabilities $\tilde{p}_i$ were defined in
\eqref{effectiveProbs}.
In this section we aim to demonstrate the characteristics of the
different measurements we have defined, and the relations between
them, via the solutions for a specific ensemble.

The optimal measurement for each criterion was computed by explicitly
solving the optimization problems \eqref{aimProb} and \eqref{AnalProb}
for the relevant cases.
The computation was done using the SeDuMi toolbox in Matlab.

\subsection{A Three-State System with Uniform Uncertainty}
\label{numB}

We examine an ensemble comprised of the pure states
$\rho_i^0 = \ket{u_i} \bra{u_i}$ in a three dimensional Hilbert space,
\begin{equation} \label{3in3}
u_1 = \frac{1}{\sqrt{5}} \begin{pmatrix} 1 \\  0 \\ -2 \end{pmatrix} \;\:
u_2 = \frac{1}{\sqrt{5}} \begin{pmatrix} 2 \\  i \\  0 \end{pmatrix} \;\:
u_3 = \frac{1}{\sqrt{2}} \begin{pmatrix} 0 \\ -1 \\  1 \end{pmatrix}
\end{equation}
with prior probabilities
\begin{equation}
p_1 = 0.2 \qquad p_2 = 0.3 \qquad p_3 = 0.5
\end{equation}
under conditions of uniform uncertainty ($q_i = q$).

\begin{figure*}
\centering
\subfigure[]{\includegraphics[width=0.32\textwidth]{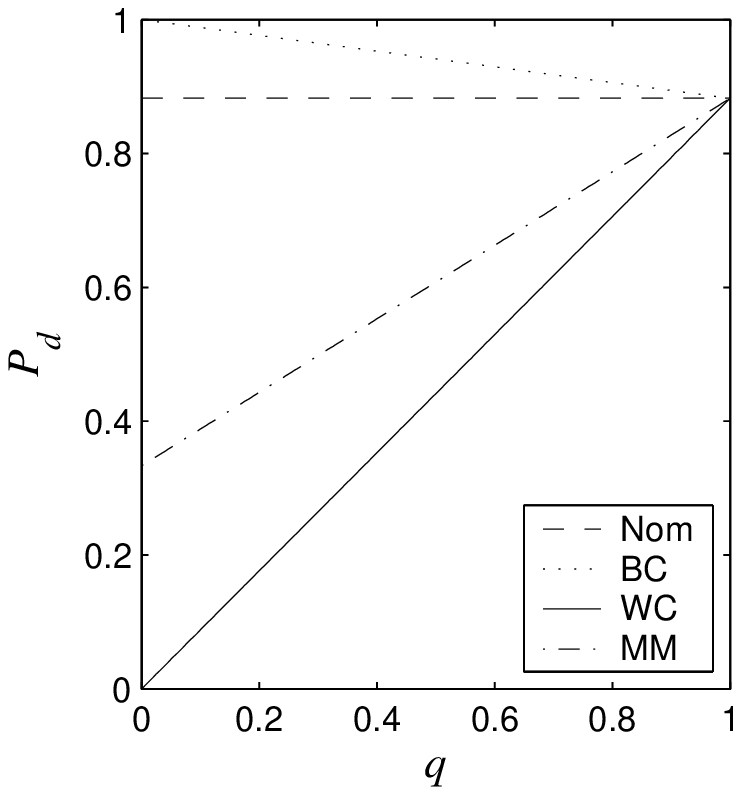}}
\subfigure[]{\includegraphics[width=0.32\textwidth]{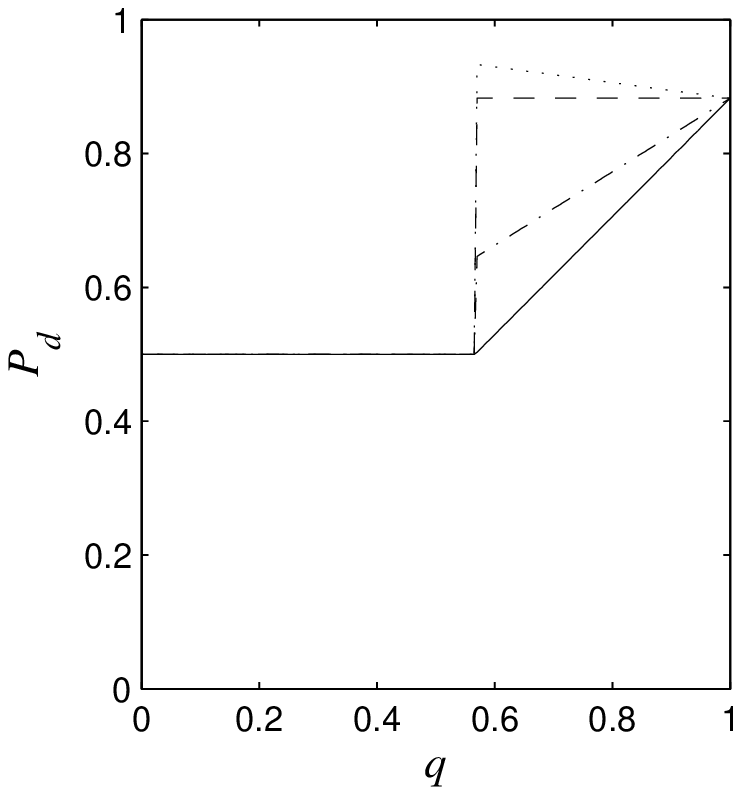}}
\subfigure[]{\includegraphics[width=0.32\textwidth]{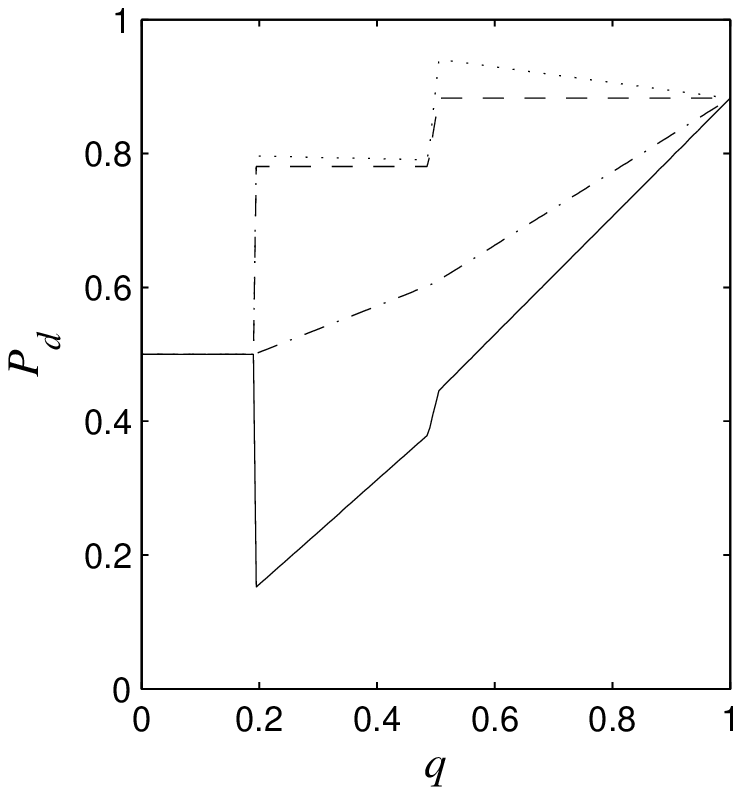}}
\caption{Probabilities of correct detection using the different
         optimal measurements as a function of $q$, for the ensemble defined in \eqref{3in3}:
         (a) optimal nominal measurement; 
         (b) optimal worst-case measurement; 
         (c) optimal average measurement. 
         The different lines in each graph are different
         manifestations of the unknown states $\rho_i^1$, where the
         abbreviations are defined in the text. \label{Detectors}}
\end{figure*}

Figure \ref{Detectors} shows the probabilities of correct detection
using the different measurements as a function of $q$.
The results referred to as `BC' and `WC' stand for `best-case' and
`worst-case' respectively \footnote{The states which achieve these
results are measurement specific, e.g. for a given value of $q$, the
unknown states $\rho_i^1$, which would generate the best result using
the nominal measurement, are not necessarily the ones that would do so
for the worst-case measurement.}.
Also shown are the results for the nominal states, that is $\rho_i^1 = \rho_i^0$,
and the results for $\rho_i^1 = \frac{1}{n} I$ (denoted `MM').

The optimal nominal measurement $\Pi^{\rm Nom}$ is independent of $q$.
Therefore the corresponding probabilities of correct detection, which
are given by expressions of the form
\begin{equation}
q P_d^{\rm Nom} +(1-q) P \big(\Pi^{\rm Nom}, \rho_i^1 (\Pi^{\rm Nom}) \big)
\end{equation}
behave linearly in $q$.

As expected from Corollary \ref{UniformWC}, the optimal worst-case
measurement $\Pi^{\rm WC}$ shows two distinct regions of behaviour.
One can verify the result by noting that for $q \leq p_{\max} / P_d^{\rm Nom}$
the optimal probability of correct detection is $P_d (q) = p_{\max}$
regardless of the choice of $\rho_i^1$ (in particular the worst and
best cases are equal), whereas for $q > p_{\max} / P_d^{\rm Nom}$ the
probabilities coincide with those obtained by $\Pi^{\rm Nom}$.

The rightmost plot in Figure \ref{Detectors} shows the results obtained
using the optimal average measurement $\Pi^{\rm Avg}$.
For high uncertainty (low $q$) this measurement also does not improve
on the lower bound $p_{\max}$.
Using the same argument as above, we conclude that in this region of $q$
$\Pi^{\rm Avg}=\bar{\Pi}$.
Note that contrary to $\Pi^{\rm WC}$, the lower bound measurement does not
appear explicitly in solution to the problem of optimal average detection.

For optimal average detection, the transition to $\Pi^{\rm Avg}=\bar{\Pi}$
occurs at a lower value of $q$ compared to $\Pi^{\rm WC}$.
An interpretation is that $\Pi^{\rm Avg}$ is a less pessimistic measurement,
relying on the input under conditions of uncertainty where $\Pi^{\rm WC}$
already regresses to guessing.

An important feature is that when $q$ is high enough so that $\Pi^{\rm Avg}$
begins to be dependant on the states themselves (and not only on the
prior probabilities), although the objective $P_d^{\rm Avg}$ improves
monotonically, the worst-case probability of correct detection does not.
In particular, there is a region of $q$ where $P_d^{\rm WC} < p_{\max}$.

This illustrates the fact that the results of each measurement are highly
dependant on the location of $\rho_i^1$ in the region of uncertainty.
The optimal average solution may have a very bad worst-case, and the
optimal worst-case solution may lead to poor detection on average.
In general, the designer of a specific setup must make similar
calculations and use cost-benefit considerations in order to choose
between the available `optimal' options.
Hence, the choice of measurement that will be eventually used is
dependant on the specific application at hand (the intended use of
the apparatus, the known states $\rho_i^0$ and probabilities $p_i$,
and the mixing bounds $q_i$).

Further insight can be revealed by examining the distances between
measurements.
Denoting $\Pi~=~\big[ \Pi_1 \;\; \Pi_2 \; \cdots \; \Pi_m \big]$
(the matrix whose columns are the measurement operators), we use the
Frobenius norm to define the \emph{measurement distance}
\begin{equation} \label{distDef}
D (\Pi,\Psi) = \| \Pi - \Psi \|_F \\
\end{equation}
and the \emph{measurement difference}
\begin{equation} \label{diffDef}
\delta \Pi (q) = \| \Pi(q) - \Pi(q - \delta q) \|_F
\end{equation}
The measurement difference can be thought of
as a gradient of the distance from a reference measurement.

\begin{figure}
\centering
\subfigure[]{\includegraphics[width=0.48\columnwidth]{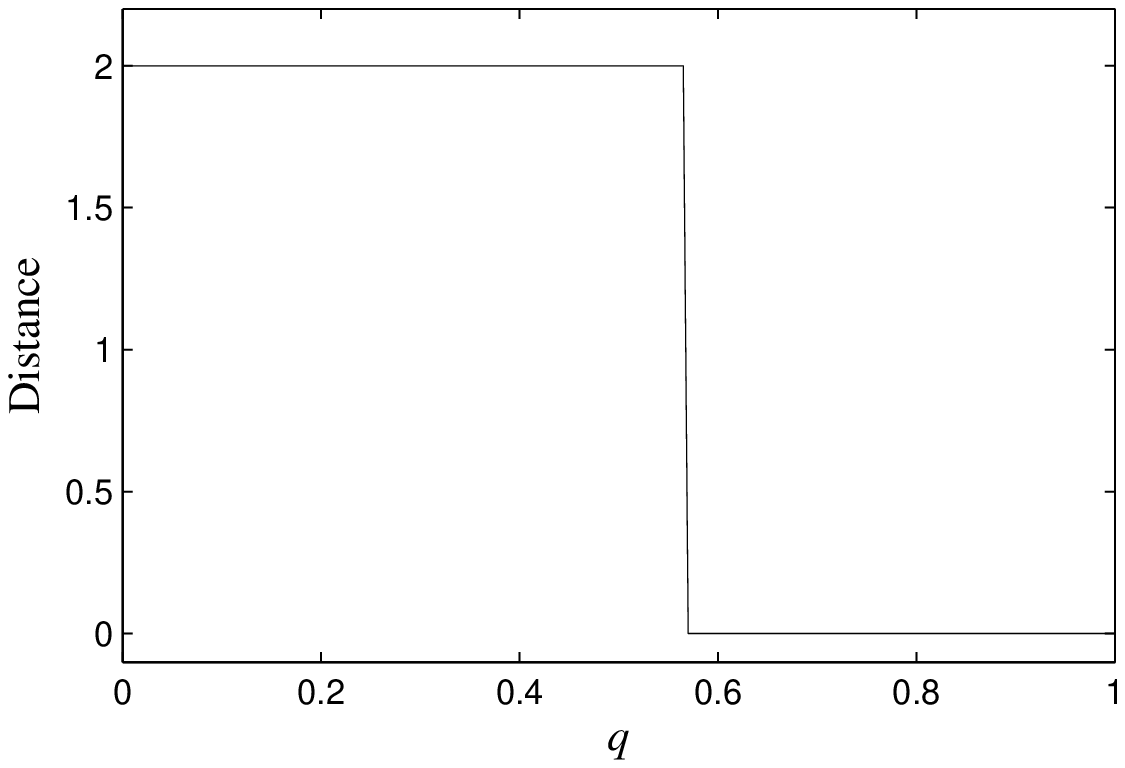}}
\addtocounter{subfigure}{1}
\subfigure[]{\includegraphics[width=0.48\columnwidth]{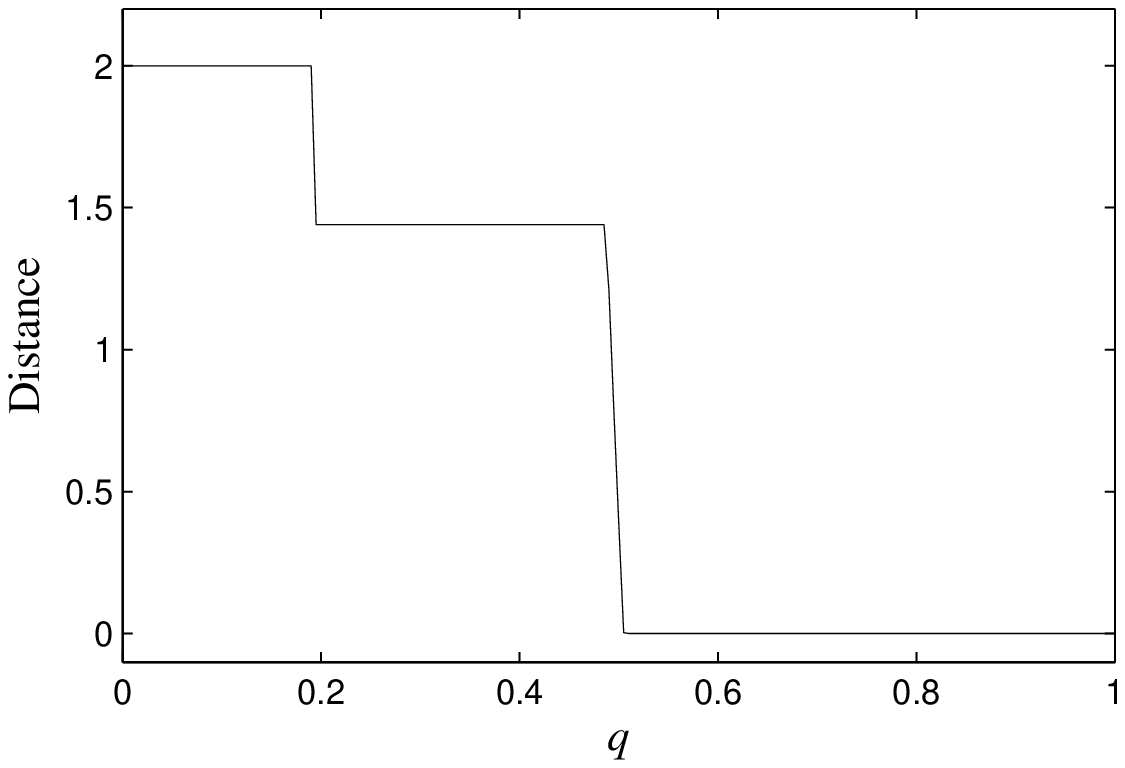}}
\addtocounter{subfigure}{-2}
\subfigure[]{\includegraphics[width=0.48\columnwidth]{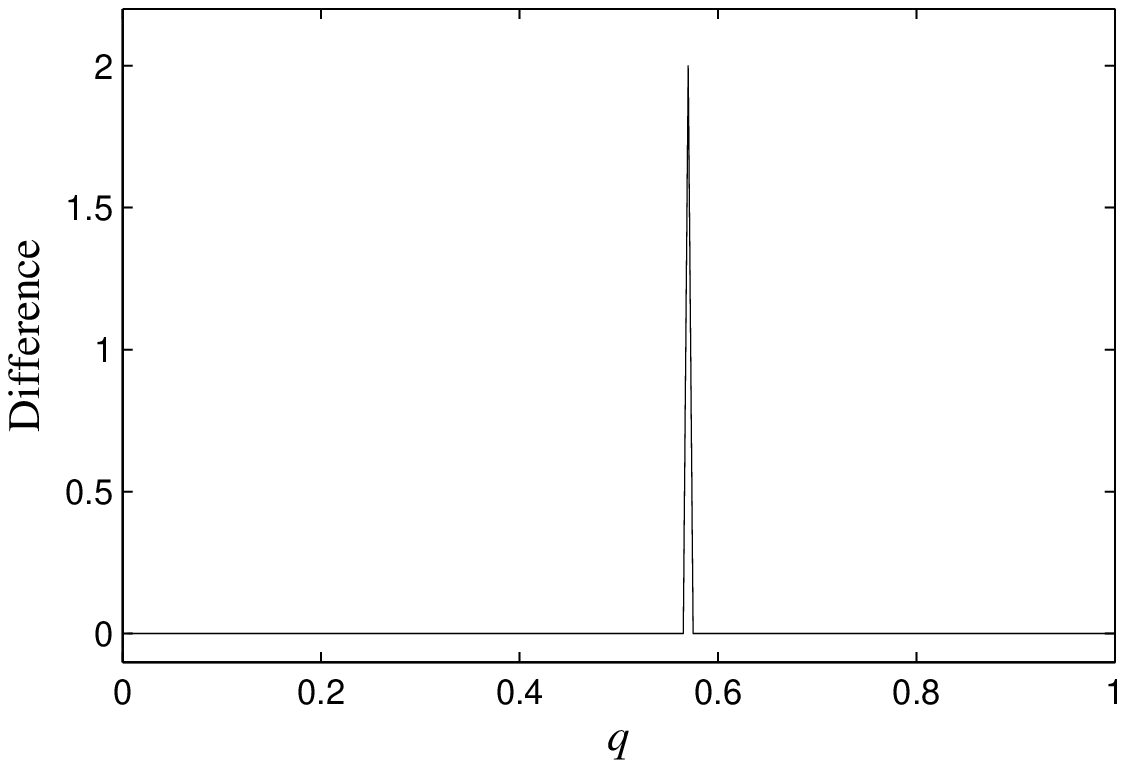}}
\addtocounter{subfigure}{1}
\subfigure[]{\includegraphics[width=0.48\columnwidth]{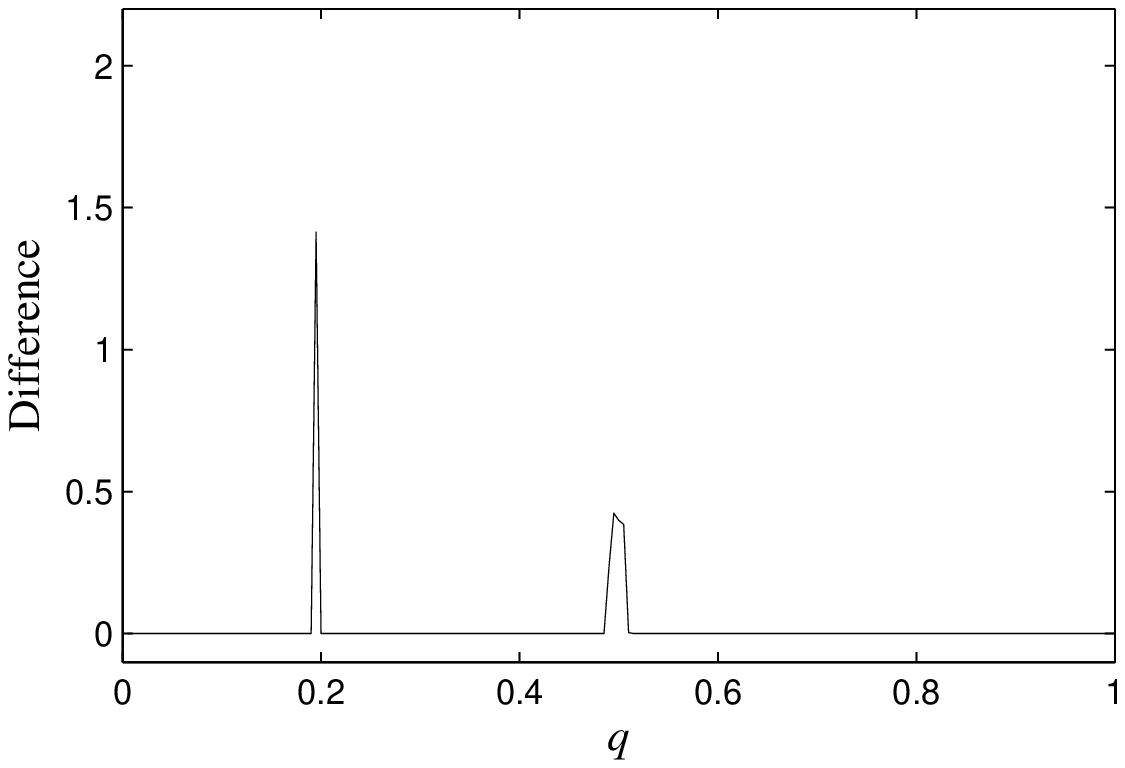}}
\caption{Measurement distances and differences as a function of $q$,
         for the ensemble defined in \eqref{3in3}:
         (a) distance between $\Pi^{\rm WC}$ and $\Pi^{\rm Nom}$;
         (b) change in $\Pi^{\rm WC}$;
         (c) distance between $\Pi^{\rm Avg}$ and $\Pi^{\rm Nom}$;
         (d) change in $\Pi^{\rm Avg}$. \label{DetDist}}.
\end{figure}

The left side of Figure \ref{DetDist} shows $D(\Pi^{\rm WC},\Pi^{\rm Nom})$
and $\delta \Pi^{\rm WC} (q)$.
The step size used in the calculations is $\delta q = 0.005$.
Because $\Pi^{\rm Nom}$ is not a function of $q$, any change in the
distance is due to change in $\Pi^{\rm WC}$.
One can clearly see the discontinuous change between the region
$q \leq p_{\max} / P_d^{\rm Nom}$, where $\Pi^{\rm WC} = \bar{\Pi}$,
and the region of low uncertainty where $\Pi^{\rm WC} = \Pi^{\rm Nom}$.
The rightmost plots show $D(\Pi^{\rm Avg},\Pi^{\rm Nom})$ and
$\delta \Pi^{\rm Avg} (q)$.
One can see that $\Pi^{\rm Avg}$ exhibits both continuous and
discontinuous change.
Moreover, in this example, for high values of $q$, we find that
$D(\Pi^{\rm Avg},\Pi^{\rm Nom})=0$.
This is characteristic of many ensembles, although as stated above,
is not universal.

\section{Conclusion}

We considered the discrimination of quantum states, drawn from a finite
set with known prior probabilities, where the states themselves are not
entirely known.
We derived two sets of necessary and sufficient conditions for the optimality
of a quantum measurement in discriminating between the states.
The first in the sense of minimal worst-case probability of a detection error,
the second in the sense of minimal average probability of detection error.
In both cases, the uncertainty is manifested as optimally discriminating among
an ``effective ensemble''.
We found that under our model, when the uncertainty is of uniform magnitude
for all states, one can, in many cases, ignore it altogether.

Possible avenues for further work are looking into structured uncertainty -
where the unknown states $\rho_i^1$ are known to some extent (for example
their range-space is restricted to a certain subspace), and leakage between
the possible states, i.e. the unknown states are a mixture of the known states
$\rho_i^1 = \sum_j h_j \rho_j^0$.

\appendix
\section{KKT Conditions for the Worst-Case Problem}
\label{pndx1}

The Lagrangian \cite{Boyd&Vanden} of problem \eqref{AnalProb} is
\begin{equation} \label{Lagrangian}
\begin{split}
L =& -\sum_{i=1}^m p_i \Tr \left[ \left( q \Pi_i +
 (1-q) \lambda_i I \right) \rho_i^0 \right] - \sum_{i=1}^m w_i \lambda_i \\
 &- \sum_{i=1}^m \Tr \left[ Z_i (\Pi_i- \lambda_i I) \right]
 + \Tr \biggl[ U (I - \sum_{i=1}^m \Pi_i) \biggr] 
\end{split}
\end{equation}
where $w_i \geq 0 $, $Z_i \geq 0$ and $U$ are the Lagrange multipliers.

The KKT necessary and sufficient conditions for optimality are given by
\begin{gather}
\sum_{i=1}^m \Pi_i = I \qquad \quad  \Pi_i \geq \lambda_i I \geq 0 \label{KKT1} \\
\sumhight Z_i \geq 0 \qquad \qquad w_i \geq 0 \label{KKT2} \\
\frac{\partial L}{\partial \Pi_i} = -p_i q_i \rho_i^0 - Z_i + U = 0 \label{KKT3} \\
\frac{\partial L}{\partial \lambda_i} = -p_i (1-q_i) +\Tr(Z_i) - w_i = 0 \label{KKT4} \\
\sumhight Z_i (\lambda_i I - \Pi_i ) = 0 \label{KKT5} \\
w_i \lambda_i = 0
\end{gather}
From \eqref{KKT3} and \eqref{KKT4} we find that at the optimum
\begin{align}
Z_i &= U - p_i q_i \rho_i^0 \\
w_i &= \Tr(U)-p_i
\end{align}
Using these relations, the KKT conditions can be recast in the
form \eqref{N&S1}.

\section{Proof of Lemma 1}
\label{pndx2}

Given a random pure state $\ket{\mu}$, with uniform distribution
over the unit sphere, and an arbitrary orthogonal basis
$\{\ket{\pi_k}\}_{k=1}^n$, the $n$ probabilities
\begin{equation}
\sigma_k \triangleq | \bracket{\pi_k}{\mu} |^2
\end{equation}
form a random vector in the $n-1$ dimensional simplex $B$,
defined by the conditions $\sigma_k \geq 0$ for all $k$ and
$\sum_k \sigma_k = 1$.
The distribution of $\vec{\sigma}$ is given by (see \cite{Wootters1,Sykora1})
\begin{equation}
\sigma_k = \frac{y_k}{\sum_{k=1}^n y_k}
\end{equation}
where $\{y_k\}_{k=1}^n$ are independent random variables
obeying an exponential distribution with parameter $1$.

Due to the fact that $B$, the domain in which $\vec{\sigma}$ is
distributed, is a convex set, the expectation value of $\vec{\sigma}$
must lie in $B$, i.e. $\langle \vec{\sigma} \rangle \in B$.
The distribution of $\vec{\sigma}$ is, of course, symmetrical with
respect to exchange of any of its coordinates
($\sigma_k \leftrightarrow \sigma_{k'}$), and then so must be
the average.
The only such symmetrical point in $B$ is
$(\frac{1}{n},...,\frac{1}{n})$.
And so
\begin{equation} \label{ExpectedProbability}
\langle \sigma_k \rangle = \frac{1}{n}
\end{equation}

Given a quantum state that has undergone a random rotation $U$
\begin{equation}
\rho = UGU^*
\end{equation}
we can assume without loss of generality that the
state $G$ is diagonal, thereby permitting us to rewrite $\rho$
in the form
\begin{equation} \label{CombPurePndx}
\rho = \sum_{j=1}^n g_j \ket{\mu_j} \bra{\mu_j}
\end{equation}
with $\sum_{j=1}^n g_j = 1$ and
$\bracket{\mu_j}{\mu_{j'}} = \delta_{j,j'}$.
We now express $\ket{\mu_j}$ using their harmonic expansions in
the eigenvectors of $\Pi$.
\begin{equation} \label{expansion}
\ket{\mu_j} = \sum_{k=1}^n \alpha_{jk} \ket{\pi_k}
\end{equation}
where $\ket{\pi_k}$ is the eigenvector of $\Pi$
corresponding to eigenvalue $\lambda_k$.
Substituting \eqref{expansion} in \eqref{CombPurePndx} we get
\begin{equation}
\begin{split}
\rho &= \sum_{j=1}^n g_j \sum_{k=1}^n \alpha_{jk}
      \ket{\pi_k} \sum_{k'=1}^n \alpha_{jk'}^* \bra{\pi_{k'}} \\
   &= \sum_{j=1}^n \sum_{k,k'=1}^n g_{j} \alpha_{jk}
      \alpha_{jk'}^* \ket{\pi_k} \bra{\pi_{k'}}
\end{split}
\end{equation}
and
\begin{equation}
\begin{split}
\Tr(\Pi \rho) &= \sum_{j=1}^n \sum_{k,k'=1}^n g_j
   \alpha_{jk} \alpha_{jk'}^* \Tr(\Pi \ket{\pi_k} \bra{\pi_{k'}}) \\
&= \sum_{j=1}^n \sum_{k,k'=1}^n g_j \alpha_{jk} \alpha_{jk'}^*
   \lambda_{k} \underset{\delta_{k,k'}}{\underbrace{\bracket{\pi_{k'}}{\pi_k}}} \\
&= \sum_{j=1}^n \sum_{k=1}^n g_j |\alpha_{jk}|^2 \lambda_k
\end{split}
\end{equation}

Although $\ket{\mu_j}$ are not probabilistically independent (they
are mutually orthogonal), their marginal distributions are all uniform.
The expectation values of the squared modulus are (from
\eqref{ExpectedProbability})
\begin{equation}
\langle | \alpha_{jk} |^2 \rangle =
      \langle | \bracket{\pi_k}{\mu_j} |^2 \rangle = \frac{1}{n}
\end{equation}
which in turn leads to
\begin{equation}
\begin{split}
\langle \Tr(\Pi \rho) \rangle &= \sum_{j=1}^n \sum_{k=1}^n
  \frac{g_j \lambda_k}{n} \\
&= \frac{1}{n} \sum_{j=1}^n g_j \sum_{k=1}^n \lambda_k = \frac{1}{n} \Tr(\Pi)
\qedhere
\end{split}
\end{equation}


\begin{acknowledgments}
We would like to thank Netanel Lindner and Ami Wiesel, for fruitful discussions concerning this work.
\end{acknowledgments}

\bibliography{UnmodelledUncertaintyBIB}

\end{document}